\documentclass{llncs}
\pdfoutput=1
\usepackage{hyperref}
\usepackage{ifpdf}
\ifpdf
 \usepackage[pdftex]{graphicx}
\else
 \usepackage[dvips]{graphicx}
\fi
\usepackage{color}
\usepackage{url}
\usepackage{wrapfig}

\newcommand{\WGL}{{\em WGL}}

\title{The Web Geometry Laboratory Project\thanks{The final
    publication is available at http://link.springer.com.}}

\author{Pedro Quaresma\inst{1} 
  \and Vanda Santos\inst{2} 
  \and Seifeddine Bouallegue\inst{3}\thanks{IAESTE traineeship PT/2012/71.}
} 

\institute{
  CISUC/Department of Mathematics, University of Coimbra \\
  3001-454 Coimbra, Portugal, \email{pedro@mat.uc.pt} \\
  \and
  CISUC, 3001-454 Coimbra, Portugal, \email{vsantos7@gmail.com} \\
  \and Innov'Com / University of Carthage, Tunisia,
  \email{saief.bouallegue@gmail.com} } \date{}

\begin{document}

\maketitle

\begin{abstract}
  The \emph{Web Geometry Laboratory} (\WGL) project's goal is to build
  an adaptive and collaborative blended-learning Web-environment for
  geometry.

  In its current version (1.0) the WGL is already a collaborative
  blended-learning Web-environment integrating a dynamic geometry
  system (DGS) and having some adaptive features. All the base
  features needed to implement the adaptive module and to allow the
  integration of a geometry automated theorem prover (GATP) are also
  already implemented.

  The actual testing of the WGL platform by high-school teachers is
  underway and a field-test with high-school students is being
  prepared.

  The adaptive module and the GATP integration will be the next steps
  of this project.

\end{abstract}

\section{Introduction}
\label{sec:introduction}

The use of intelligent computational tools in a learning environment
can greatly enhance its dynamic, adaptive and collaborative
features. It could also extend the learning environment from the
classroom to outside of the fixed walls of the school.

To build an adaptive and collaborative blended-learning environment
for geometry, we claim that we should integrate dynamic geometry
systems (DGSs), geometry automated theorem provers (GATPs) and
repositories of geometric problems (RGPs) in a Web system capable of
individualised access and asynchronous and synchronous interactions. A
system with that level of integration will allow building an
environment where each student can have a broad experimental learning
platform, but with a strong formal support. In the next paragraphs we
will briefly explain what do we mean by each of these features and how
the Web Geometry Laboratory (WGL) system cope, or will cope, with
that.

\paragraph{A blended-learning environment} is a mixing of different
learning environments, combining traditional face-to-face classroom
(synchronous) methods with more modern computer-mediated
(asynchronous) activities. A Web-environment is appropriate for both
situations (see Figure~\ref{fig:globalnet}).

\paragraph{An adaptive environment} is an environment that is able to
adapt its behaviour to individual users based on information acquired
about its user(s) and its environment and also, an important feature
in a learning environment, to adapt the learning path to the different
users needs. In the \WGL\ project this will be realised through the
registration of the geometric information of the different actions
made by the users and through the analysis of those
interactions~\cite{Quaresma2012b}.

\paragraph{A collaborative environment} is an environment that allows
the knowledge to emerge and appear through the interaction between its
users. In WGL this is allowed by the integration of a DGS and by the
users/groups/constructions relationships.

Using a DGS, the constructions are made from free objects
and constructed objects using a finite set of property preserving
manipulations. These property preserving manipulations allow the
development of ``visual proofs'', these are not formal proofs. The
integration in the WGL of a GATP will give its users the possibility
to reason about a given DGS construction, this is an actual formal
proof, eventually in a readable format. They can be also used to test
the soundness of the constructions made by a
DGS~\cite{Janicic2010,Janicic2007}.

As said above to have an adaptive and collaborative blended-learning
environment for geometry we should integrate intelligent geometric
tools in a Web system capable of asynchronous and synchronous
interactions. This integration is still to be done, there are
already many excellent DGSs~\cite{wikipedia2013}, some of them have
some sort of integration with GATPs, others with
RGP~\cite{Janicic2010,Quaresma06d}. Some attempts to integrate these
tools in a learning management system (LMS) have already been done,
but, as far as we know, all these integrations are only partial
integrations. A learning environment where all these tools are
integrated and can be used in a fruitful fashion does not exist
yet~\cite{Santos2013}.

\section{The Web Geometry Laboratory Framework}
\label{sec:CollaborativeEnvironmentForGeometry}

\begin{wrapfigure}[10]{r}{4cm} 
  \includegraphics[width=4cm]{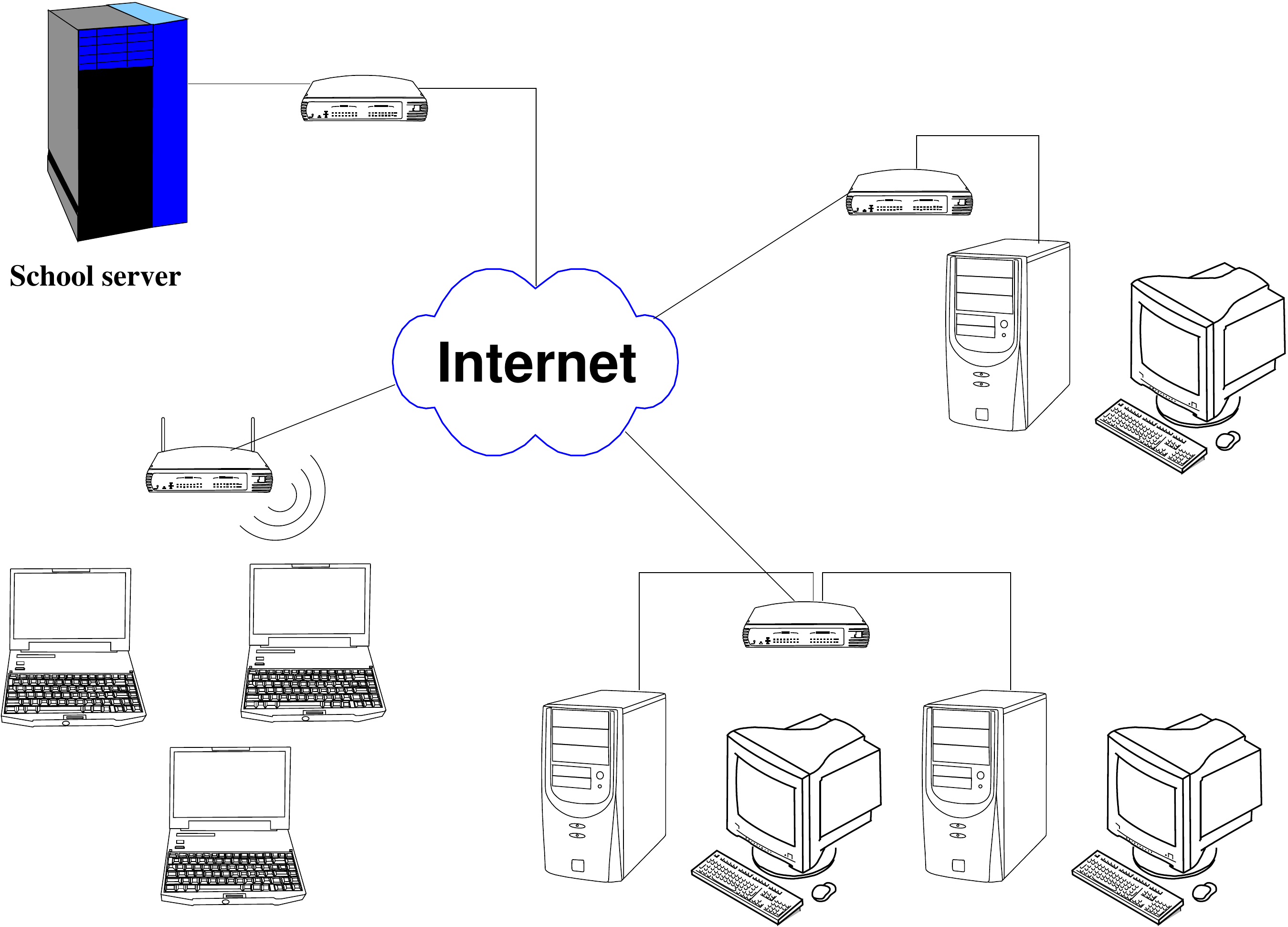}
  \caption{School Server}
  \label{fig:globalnet}
\end{wrapfigure}

A class session using \WGL\ is understood as a Web laboratory where
all the students (eventually in small groups) and the professor will
have a computer running \WGL\ clients. Also needed is a \WGL\ server,
e.g. in a school Web-server (see Figure~\ref{fig:globalnet}).

The \WGL\ server is the place where all the information is kept: the
login information; the group definition; the geometric constructions
of each user; the users activity registry; etc. In the \WGL\ server is
also kept the DGS applet and the GATP will also execute there. Each
client will have an instance of the DGS applet, using the server to
all the needed information exchange.

After installing a \WGL\ server the administrator of the system should
define all the teachers that will be using the system. The teachers
will be privileged users in the sense that they will be capable of
define other users, their students. In the beginning of each school
year the teachers will define all his/her students as regular users of
the \WGL. The teacher may also define groups of users (students),
these groups can be define at any given time, e.g. for a specific
class, and it will be within this groups that the collaboration
between its members will be possible. The definition of the groups and
the membership relation between groups and its members will be the
responsibility of the teachers that could create groups, delete groups
and/or modify the membership relation at any given time.

Each user will have a ``space'' in the server where he/she can keep
all the geometric construction that he/she produces. Each user will
have full control over this personal scrapbook, having the possibility
of saving, modifying and deleting each and every construction he/she
produces using the DGS applet.

To allow the collaborative work a permissions system was
implemented. This system is similar to the ``traditional Unix
permissions'' system. The users will own the geometric construction
defining the {\bf r}eading, {\bf w}riting and {\bf v}isibility
permissions (rwv) per geometric construction. The users to groups and
the constructions to groups relationships can be established in such a
way that the collaborative working, group-wise, is possible.

By default, the teacher will own all the groups he/she had created
granting him/her, in this way, access to all the constructions made by
the students. The default setting will be {\tt rwvr-v---}, meaning
that the creator (owner) will have all the permissions, other users
belonging to his/her groups will have ``read'' access and all the
others users will have none. At any given moment he/she can download
(read) the construction into the DGS, modify it and, eventually,
upload the modified version into the database.

\begin{figure}[hbtp]  
  \centering{
    \includegraphics[width=1\textwidth]{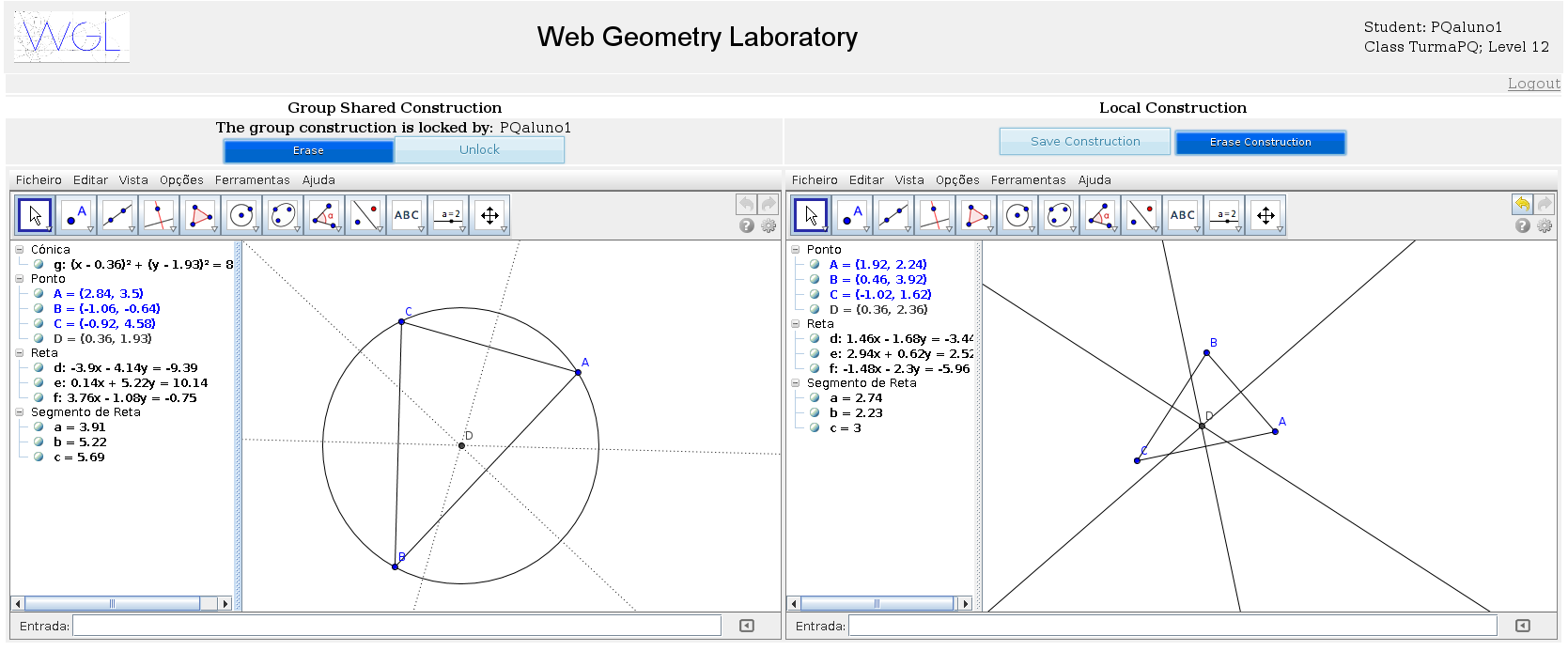}
  }
  \caption{Students' Interface}
  \label{fig:withlock}
\end{figure}

The collaborative module of \WGL\ distinguishes students having the
lock over the group construction from those without the lock. The
students with the lock will have a full-fledged DGS applet, and they
will be working with the group construction (see
Figure~\ref{fig:withlock}). The students without the lock will have
also the two DGS applets, but the construction in the ``group shared
construction'' one is a synchronised version of the one being
developed by the student with the lock, and a full-fledged version
that can be used to develop his/her own efforts. A text-chat will be
available to exchange information between group members. The teacher
could always participate in this efforts having for that purpose an
interface where he/she can follow the students and groups activities.

The \WGL\ collaborative features are thought mostly for a
blended-learning setting, that is, a classroom/laboratory where the
computer-mediated activities are combined with a face-to-face
classroom interaction. Nevertheless given the fact that the \WGL\ is a
Web application the collaborative work can extend itself to the
outside of the classroom and be used to develop collaborative work at
home, e.g. solving a given homework. In this setting the only drawback
it will be a slow connection to the \WGL\ server. We estimate that a
normal bandwidth ($\geq 20$Mb) will be enough.

The \WGL\ as a Web client/server application; the database (to keep:
constructions; users information; constructions; permissions; user's
logs); the DGS applet; the GATP and the synchronous and asynchronous
interaction are all implemented using free cross-platform software,
namely PHP, Javascript, Java, AJAX, JQuery and MySQL, and also
Web-standards like XHTML, CSS style-sheets and XML. The WGL is a
internationalised tool (i18n/l10n) with already translations for
Portuguese and Serbian, apart from default support for English. All
this will allow to build a collaborative learning environment where
the capabilities of tools such as the DGS and the GATP can be used in
a more rich setting that in an isolated environment where (eventually)
every students could have a computer with a DGS but where the
communication between them would be non-existent. The exchange of
text, oral and geometric information between members of a group will
enrich the learning environment.

Learning environments supported by computer are seen as an important
means for distance education.  The DGS are also important in classroom
environments, as a much enhanced substitute for the ruler and compass
physical instruments, allowing the development of experiments,
stimulating learning by experience. There are several DGS available,
such as: {\em GeoGebra}, {\em Cinderella}, {\em Geometric Supposer},
{\em GeometerSketchpad}, {\em CaR}, {\em Cabri}, {\em GCLC} but none
of then defines a Web learning environment with adaptive and
collaborative features~\cite{Santos2013}.  The program {\em
  Tabul{\ae}} is a DGS with Web access and with collaborative
features. This system is close to \WGL, the permissions system and
the fact that the DGS is not ``hardwired'' to the system but it is an
external tool incorporated into the system, are features that
distinguish positively \WGL\ from {\em Tabul{\ae}}. The adaptive
features, the connection to the GATP and the
internationalisation/localisation are also features missing in {\em
  Tabul{\ae}}~\cite{Santos2013}.

\section{Conclusions and Future Work}
\label{sec:ConclusionsFutureWork}

When we consider a computer system for an educational setting in
geometry, we feel that a collaborative, adaptive blended-learning
environment with DGS and GATP integration is the most interesting
solution. That leads to a Web system capable of being used in the
classroom but also outside the classroom, with collaborative and
adaptive features and with a DGS and GATPs integrated.

The \WGL\ system is a work-on-progress system. It is a client/server
modular system incorporating a DGS, some adaptive features, i.e., the
individualised scrapbook where all the users can keep their own
constructions and with a collaborative module. Given the fact that it
is a client/server system the incorporation of a GATP (on the server)
it will not be difficult. One of the authors has already experience on
that type of integration~\cite{Janicic2007,Quaresma2011,Quaresma06d}.

A first case study, involving two high-schools (in the North and
Center of Portugal) three classes, two teachers and 44 students and
focusing in the use of \WGL\ in a classroom, is already being prepared
and it will be implemented in the spring term of 2013.

The next task will be the adaptive module, the logging of all the
steps made by students and teacher and the construction of student's
profiles on top of that. The last task will be the integration of the
GATP in the \WGL. We hope that at the end the \WGL\ can became an
excellent learning environment for geometry.

A prototype of the \WGL\ system is available at
\url{http://hilbert.mat.uc.pt/WebGeometryLab/}. You can enter as
``anonymous/anonymous'', a student-level user, or as
``cicm2013/cicm'', a teacher-level user.

\newcommand{\noopsort}[1]{} \newcommand{\singleletter}[1]{#1}

\end{document}